\documentclass[prl,twocolumn,floatfix,showpacs]{revtex4-1}
\usepackage{graphicx,color}
\usepackage{amsmath,amssymb,bm}

\newcommand{\tn}{\textnormal}
\newcommand{\cprb}[3]{Phys.~Rev.~B {\bf #1}, #2 (#3)}
\newcommand{\cprl}[3]{Phys.~Rev.~Lett.~{\bf #1}, #2 (#3)}

\definecolor{darkred}{rgb}{0.90,0,0}
\definecolor{darkgreen}{rgb}{0,0.60,.2}
\definecolor{darkblue}{rgb}{0,0,1}
\definecolor{grey}{cmyk}{0,0,0,0.25}
\definecolor{orange}{cmyk}{0,0.6,0.8,0}

\begin{document}
\title{\boldmath Finite temperature dynamical DMRG and the Drude weight of spin-$1/2$ chains}
\author{C.\ Karrasch$^{1}$}
\author{J.\ H.\ Bardarson$^{1,2}$}
\author{J.\ E.\ Moore$^{1,2}$}

\affiliation{$^1$Department of Physics, University of California, Berkeley, California 95720, USA}

\affiliation{$^2$Materials Sciences Division, Lawrence Berkeley National Laboratory, Berkeley, CA 94720, USA}

\begin{abstract}

We propose an easily implemented approach to study time-dependent correlation functions of one dimensional systems at finite temperature $T$ using the
density matrix renormalization group. The entanglement growth inherent to any time-dependent calculation is significantly reduced if the auxiliary degrees of freedom which purify the statistical operator are time evolved with the physical Hamiltonian but reversed time. We exploit this to investigate the long time behavior of current correlation functions of the XXZ spin-$1/2$ Heisenberg chain. This allows a direct extraction of the Drude weight $D$ at intermediate to large $T$. We find that $D$ is nonzero -- and thus transport is dissipationless -- everywhere in the gapless phase. At low temperatures we establish an upper bound to $D$ by comparing with bosonization.  

\end{abstract}

\pacs{71.27.+a, 75.10.Pq, 75.40.Mg, 05.60.Gg}
\maketitle



It is an intriguing question if a physical system can exhibit dissipationless transport. In this case the conductivity has a singular contribution 
$D\delta(\omega)$ where $D$ is typically referred to as the Drude weight. As a consequence, a fraction of an initially excited current will survive to
infinite time. If the current operator is conserved by the Hamiltonian, the corresponding quantum system clearly supports dissipationless
transport at any temperature $T$. The more subtle question of whether the Drude weight can be nonzero when the current operator has no overlap with any
local conserved quantity has attracted considerable attention
\cite{bethezotos,bethekluemper,ed1,ed2,ed3,ed4,ed5,qmc1,qmc2,sirkerJ1,sirkerJ2,integrab1,integrab2,fieldtheory,fabian,mazurnonloc,mazurnonloc2,znidaric} without being resolved. In one spatial dimension it is believed that $D\neq0$ is only possible at $T>0$ for an integrable model where an infinite set of conserved local operators might restrict dissipation processes.

This paper uses a finite-temperature time-dependent density matrix renormalization group (DMRG) approach to calculate the Drude weight for a prototypical integrable one-dimensional system -- the antiferromagnetic XXZ Heisenberg chain. The latter describes $L\to\infty$ interacting spin-$1/2$ degrees of freedom $S_n^{x,y,z}$
\begin{equation}
H = J\sum_{n=1}^{L} \left(S^x_{n}S^x_{n+1} + S^y_{n}S^y_{n+1} + \Delta S^z_{n}S^z_{n+1}\right)~, 
\end{equation}
or equivalently spinless fermions through the Jordan-Wigner transformation. The spectrum is gapless for all anisotropies $0\leq\Delta\leq1$. While Bethe ansatz allows evaluation of $D$ at $T=0$ via the Kohn formula \cite{kohn,betheT0}, it has not been possible to reliably compute its value for nonzero temperatures. The Mazur lower bound \cite{mazurbound} determined by all local conserved quantities is zero in absence of a magnetic field (however, a lower bound was recently constructed \cite{mazurnonloc,mazurnonloc2} from a nonlocal operator at $T=\infty$; we will explicitly investigate whether or not it saturates the Drude weight). Two different Bethe ansatz approaches \cite{bethezotos,bethekluemper} yield contradictory results. Exact diagonalization \cite{ed1,ed2,ed3,ed4} can only treat systems of up to $L\sim20$ sites, and quantum Monte Carlo \cite{qmc1,qmc2} data requires an ill-controlled analytic continuation in order to extract $D$. A recent bosonization study combined with DMRG numerics \cite{sirkerJ1,
sirkerJ2} proves the existence of a diffusive transport channel \cite{qmc3} and yields an upper bound on the Drude weight; but the timescales reachable within the DMRG are yet too small to make a decisive statement about whether or not $D$ is nonzero. While most of these works conclude in favor of a finite Drude weight in the gapless phase, no ultimately accepted quantitative picture (e.g., concerning its $\Delta$ and $T$ -- dependence and whether or not it vanishes for the isotropic chain) has emerged. It is the first goal of this work to fill this gap at intermediate and high temperatures.

The density matrix renormalization group \cite{dmrgrev} was initially devised as a powerful tool to calculate ground state properties of one-dimensional systems with short-ranged interactions \cite{white1,white2} and subsequently extended to address real-time dynamics \cite{tdmrg1,tdmrg2,tdmrg3,tdmrg4}. In principle, it can be applied straightforwardly to nonzero $T$ \cite{dmrgT,barthel,metts} by introducing auxiliary degrees of freedom to purify the thermal statistical operator \cite{suzuki}. In practice, however, the increase of entanglement with time has limited previous finite-temperature calculations to rather short timescales \cite{sirkerJ1,sirkerSZ,sirkerSZ2}. The second aim of our paper is to propose an easy-to-implement modification to the DMRG algorithm of Ref.~\onlinecite{barthel}: The auxiliaries are by construction inert, but they can be exposed to an arbitrary unitary transformation without involving any approximation. It turns out that the intuitive choice of time-evolving with the 
physical Hamiltonian but reversed time leads to a drastic reduction of the entanglement growth -- thus, significantly longer timescales can be reached and the Drude weight of the XXZ chain can be calculated after all.

This paper is organized as follows.  We first explain our modified DMRG method and test it for the exactly solvable case $\Delta=0$. It is then used to compute the Drude weight for a range of $\Delta$ and $T$. In a nutshell, we find that $D$ is nonzero everywhere from the non-interacting point $\Delta=0$ to the isotropic chain $\Delta=1$ and decreases with increasing temperature and $\Delta$; values in the gapped regime $\Delta>1$ are numerically consistent with $D=0$. We stress that our approach does not involve any approximation: The Drude weight can be read off directly from the long-time asymptotics of the spin current correlation function \cite{integrab2}; the system size is large enough (typically $L\sim100-250$) to be in the thermodynamic limit for each temperature at hand; and DMRG is numerically exact.

\begin{figure}[t]
\includegraphics[width=0.95\linewidth]{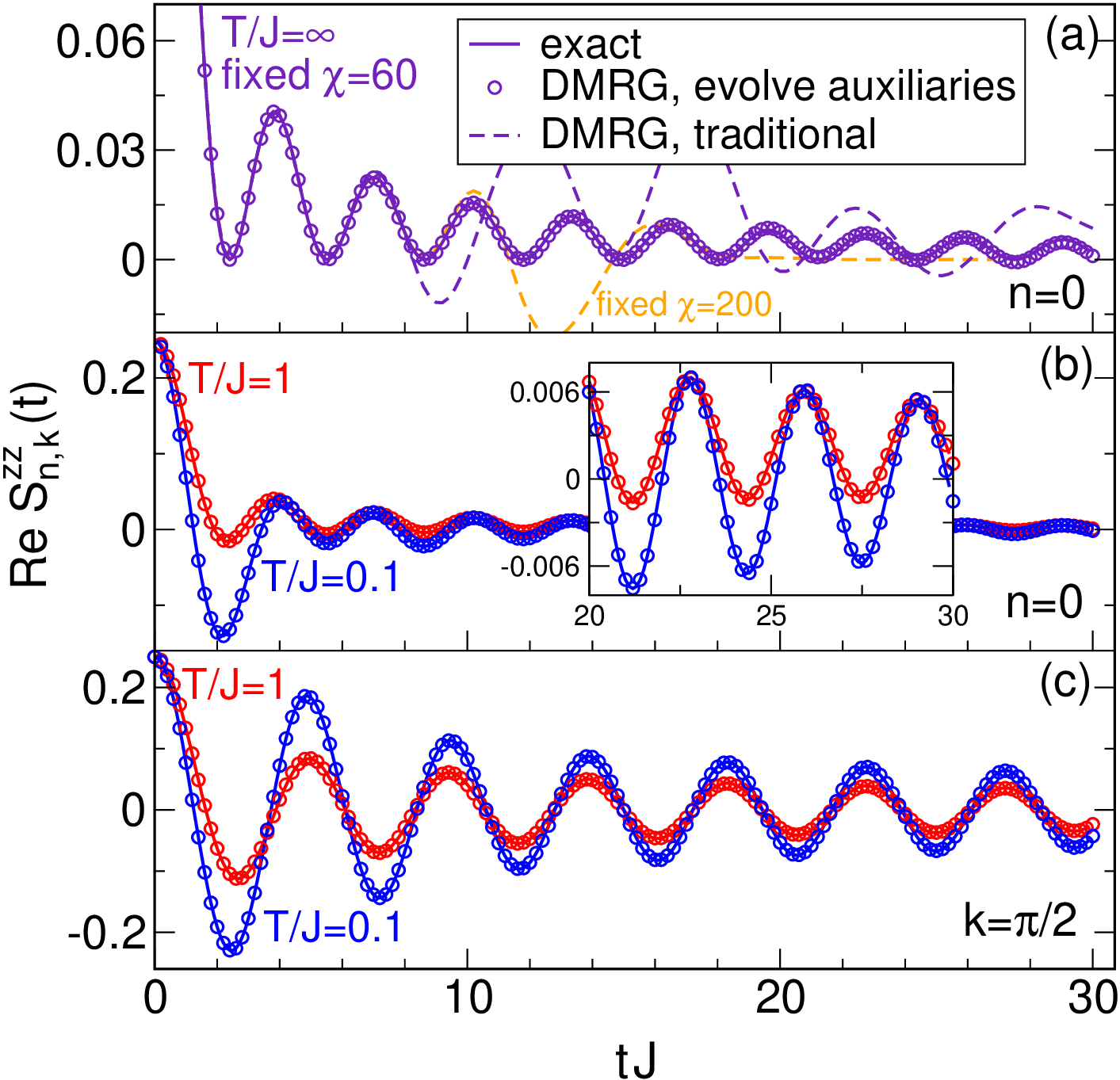}
\caption{DMRG calculation for the finite-temperature longitudinal spin structure factor of a $L=100$ site $XX$ Heisenberg chain ($\Delta=0$) in comparison with the exact result in the thermodynamic limit \cite{sirkerSZ}. We show both the real space representation $S^{zz}_n(t) = \langle S^z_{L/2+n}(t)S^z_{L/2}\rangle$ and Fourier transform $S^{zz}_k(t) = \sum_n e^{ikn} S^{zz}_n(t)$. In (a) [in (b) and (c)], the DMRG Hilbert space dimension $\chi$ [the discarded weight] is fixed. Whereas the traditional DMRG approach~\cite{barthel} breaks down rapidly, larger times can be reached if the auxiliaries are evolved with the physical Hamiltonian but reversed time. }
\label{fig:trick}
\end{figure}


\emph{Finite-temperature DMRG} --- In order to eventually compute correlation functions of the type $S^{\mu\nu}_{nm}(t) = \langle S^\mu_n(t)S^\nu_m\rangle = \tn{Tr\,}[\rho_\beta S^\mu_n(t)S^\nu_m]$ within the DMRG, one needs to purify the thermal grandcanonical density matrix \cite{suzuki} by introducing an auxiliary Hilbert space $Q$: $\rho_\beta = \tn{Tr\,}_{Q} |\Psi_\beta\rangle\langle\Psi_\beta|$. This is analytically possible only at infinite temperature $\beta=1/T=0$ where $\rho_0=2^{-L}$. However, $|\Psi_{\beta}\rangle$ can be obtained from $|\Psi_0\rangle$ by applying an imaginary time evolution, $|\Psi_\beta\rangle=e^{-\beta H/2}|\Psi_0\rangle$ \cite{dmrgrev}. Any correlation function can therefore be \textit{exactly} recast as $S^{\mu\nu}_{nm}(t)=\langle \Psi_\beta|e^{i Ht}S^\mu_n e^{-iHt}S^\nu_m |\Psi_\beta\rangle/\sqrt{\langle\Psi_\beta|\Psi_\beta\rangle}$. These objects, however, are directly accessible within a standard time-dependent DMRG framework \cite{tdmrg1,tdmrg2,tdmrg3}. It is most convenient 
to first express $|\Psi_0\rangle$ in terms of a matrix product state \cite{mps}, $|\Psi_0\rangle = \sum_{\sigma_n} A^{\sigma_1}A^{\sigma_2}\cdots A^{\sigma_{2L}}|\sigma_1\sigma_2\ldots\sigma_{2L}\rangle~,$ where $|\sigma_n\rangle$ denotes the eigenbasis of $S^z_n$, and $Q$ is spanned by the states with even indices. We choose $A^{\uparrow_\tn{odd}}=(1~0)$, $A^{\downarrow_\tn{odd}}=(0~-1)$, $A^{\uparrow_\tn{even}}=(0~1/\sqrt{2})^T$, $A^{\downarrow_\tn{even}}=(1/\sqrt{2}~0)^T$ in order to exploit $S^z$--conservation \cite{dmrgrev}. After factorizing the imaginary and real time evolutions operators $\exp(\sim H)$ using a fourth order Trotter decomposition \cite{dmrgrev}, they can be successively applied to $|\Psi_0\rangle$. At each time step, two singular value decompositions are carried out to update three consecutive matrices $A^{\sigma_n}$ (the physical sites are the odd ones and thus next-nearest neighbors). The matrix dimension $\chi$ is dynamically increased such that at each time step the sum of all squared discarded singular values is kept below a threshold value $\epsilon$. The DMRG approximation to $S^{\mu\nu}_{nm}(t)$ thus becomes exact in the limit $\epsilon\to0$.

\begin{figure}[b]
\includegraphics[width=0.95\linewidth,clip]{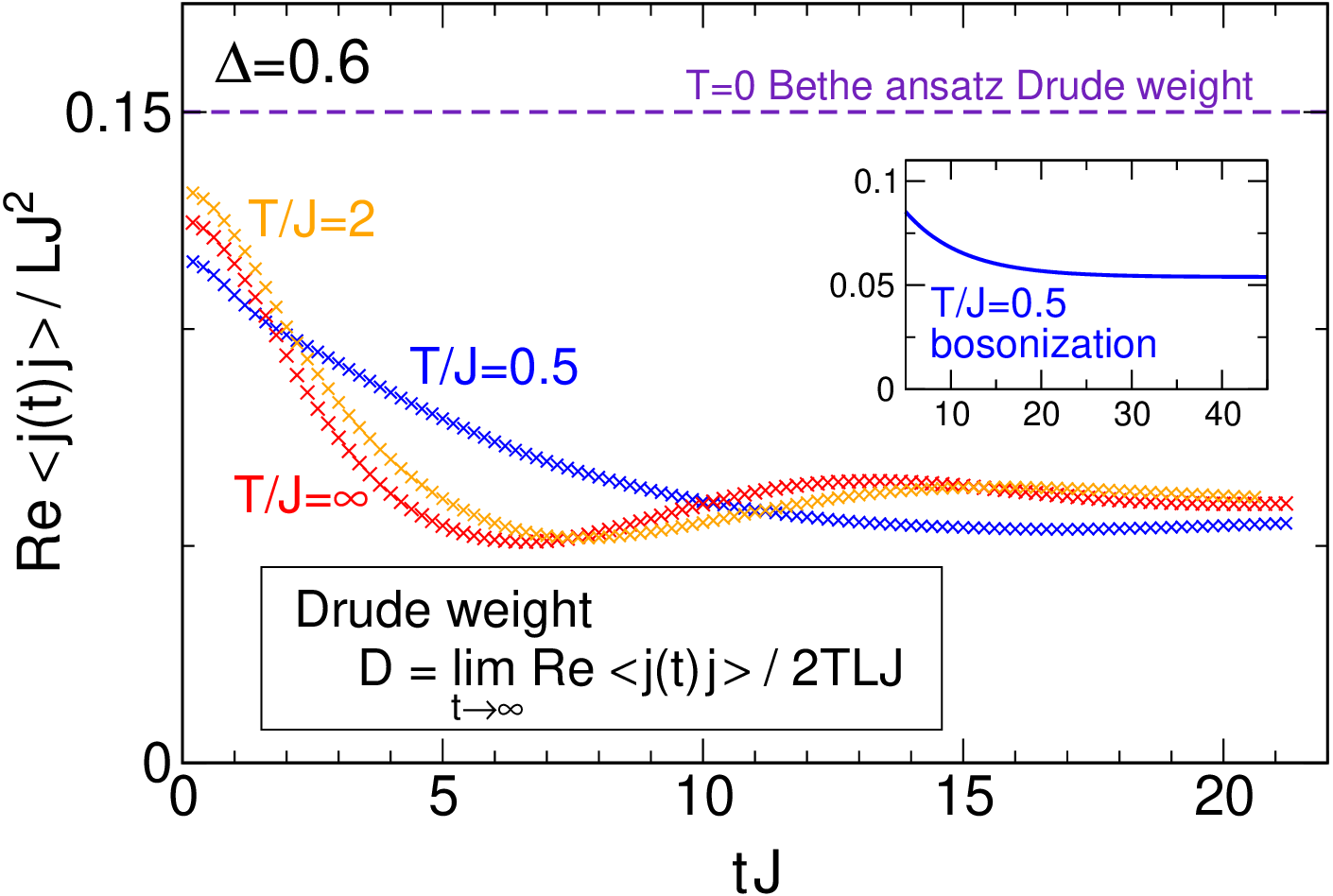}
\caption{Spin current correlation functions of the XXZ Heisenberg chain at intermediate to large temperatures. The long-time asymptotics determine the Drude weight $D$. \textit{Inset:} Bosonization result of Ref.~\onlinecite{sirkerJ2} evaluated for $D=0.05$. Despite the large temperature, bosonization agrees qualitatively with the DMRG data. }
\label{fig:highT}
\end{figure}

\begin{figure*}[t]
\includegraphics[width=0.46\linewidth,clip]{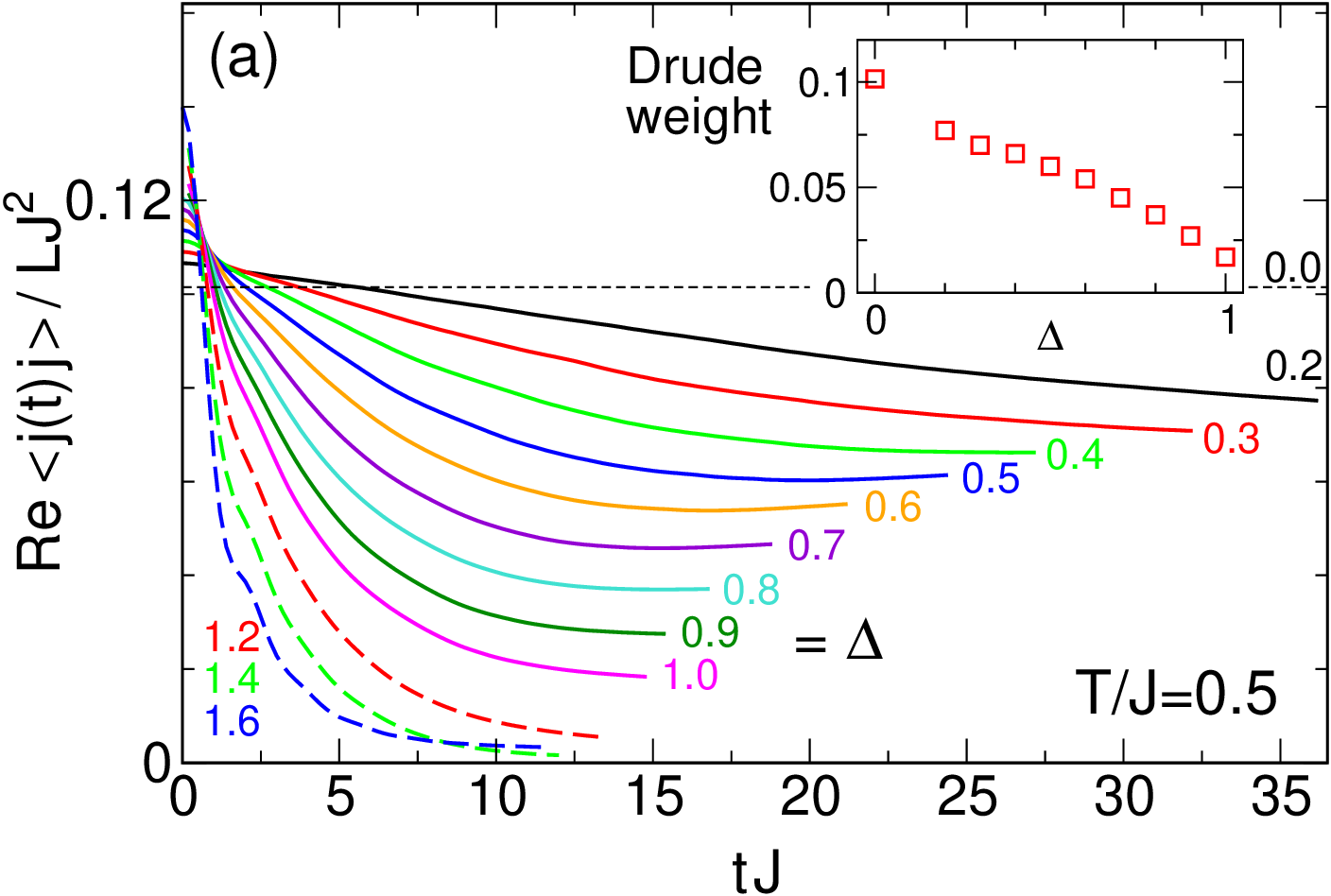}\hspace*{0.04\linewidth}
\includegraphics[width=0.46\linewidth,clip]{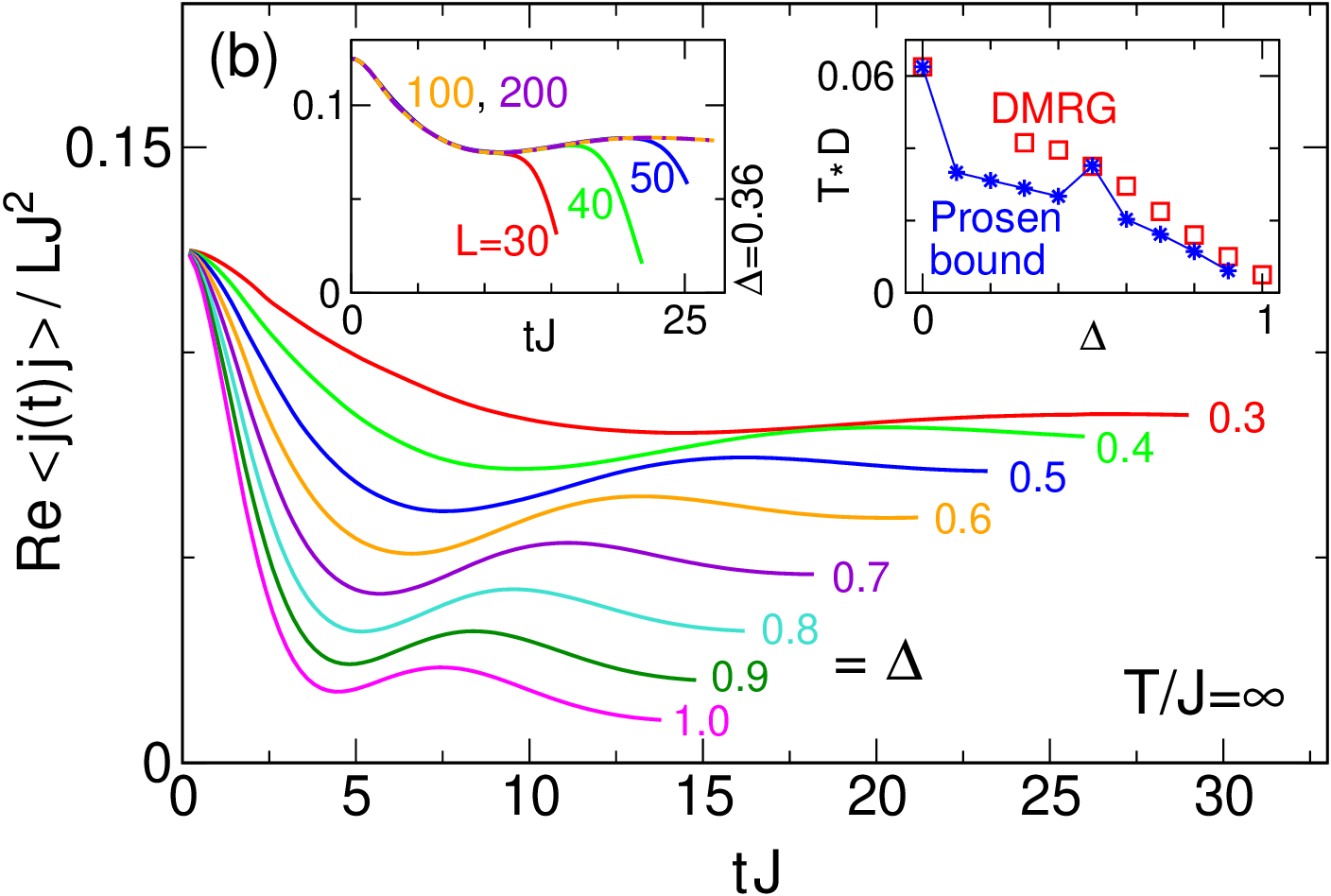}
\caption{Comprehensive picture of the $\Delta$--dependence of current correlation function at intermediate ($T/J=0.5$) and high ($T=\infty$) temperatures \cite{discweight}. The corresponding Drude weight $D$ is shown in the insets and for infinite $T$ compared to the lower bound of Ref.~\onlinecite{mazurnonloc}. The left inset to (b) shows data for various system sizes $L=30\ldots200$, illustrating that our DMRG calculation can readily reach the thermodynamic limit.}
\label{fig:dw}
\end{figure*}


\emph{Efficient DMRG approach to $T\neq0$} --- Prior finite-temperature studies employing the DMRG were in practice limited to fairly intermediate timescales due to a rapid blow-up of the block Hilbert space dimension $\chi$ \cite{sirkerSZ,barthel}. While in a ground state calculation the entanglement entropy grows \textit{locally} around a quench $S^{\mu}_n|\tn{gs}\rangle$, at $T>0$ it increases \textit{homogeneously} even when only the thermal density matrix (i.e., the state $|\Psi_\beta\rangle$) is exposed to a supposedly trivial time evolution. However, the purification is not unique -- we have the analytic freedom to apply any unitary transformation $U_\tn{aux}(t)$ to the in principle inert auxiliary sites (physical quantities are determined by the trace over $Q$ and are thus not affected by $U_\tn{aux}$). Thus, it is reasonable to investigate: Is there some unitary operator which can undo the above damage? Indeed, the intuitive choice $U_\tn{aux}(t) = e^{+i\tilde Ht}$, with $\tilde H$ being the time-
reversed physical Hamiltonian applied to the auxilliary sites renders the time evolution of $|\Psi_\beta\rangle$ trivial \cite{commentTaux}, and the computation of $e^{-iHt}S^{\mu}_n|\Psi_\beta\rangle$ is eventually only plagued by an entanglement building up around site $n$ \cite{commentchilocal}. This leads to a significantly slower increase of $\chi$, and thus longer timescales can be reached.

To demonstrate the potential of this modification -- which does not involve any approximation except for the usual truncation inherent to each DMRG update -- we calculate the longitudinal spin structure factor of the XX chain and compare with the exact solution obtained by mapping to free fermions. The result is shown in Fig.~\ref{fig:trick}. Even for times $tJ\sim30$, $\chi$ only grows to moderate $\chi\sim250$ in order to keep the total (summed) discarded weight below $\epsilon\sim10^{-6}$. This holds for the whole temperature range $T/J=0.1\ldots\infty$ and is to be contrasted with the DMRG approaches of Refs.~\onlinecite{sirkerSZ,barthel} which break down much earlier [see Fig.~\ref{fig:trick}(a) and note that increasing $\chi$ from $60$ to $200$ merely allows $tJ\sim9$ instead of $tJ\sim7$].

In order to assure that the above scenario is generic, we have also studied (i) longitudinal and transverse spin structure factors of the XX chain in presence of a magnetic field, (ii) the XXZ chain, (iii) the anisotropic spin $1$ Heisenberg chain, and (iv) the quantum Ising model and compared with available analytics or DMRG data \cite{sirkerSZ,barthel}. The reachable timescales depend on the model and correlation function under investigation -- but introducing $U_\tn{aux}$ generally reduces the effort of finite-temperature DMRG to that of a ground state calculation.


\emph{Drude weight at intermediate to large $T$} --- We will now employ our modified DMRG to extract the Drude weight of the XXZ chain in the critical phase $0\leq\Delta\leq1$. The spin current is defined from a continuity equation and reads $j=\sum_n j_n=-\frac{iJ}{2}\sum_n(S_n^+S_{n+1}^--S_{n+1}^+S_n^-)$, where $S^\pm_n=S^x_n\pm iS^y_n$ (and $n$ denote physical sites only). The long-time asymptotics of the current correlation function directly yield $D$ \cite{integrab2}:
\begin{equation}
D = \lim_{t\to\infty} \frac{\tn{Re\,}\langle j(t) j\rangle}{2LT} = \lim_{t\to\infty} \sum_n \frac{\tn{Re\,}\langle j_n(t) j_{L/2}\rangle}{2T}~.
\end{equation}
A similar strategy was pursued in Ref.~\onlinecite{sirkerJ1} using the transfer matrix DMRG. We quantitatively reproduce those results but are able to reach time regimes about a factor of $2-3$ larger within one day of computer time \cite{commentnum}. At intermediate temperatures $T/J\gtrsim0.5$, this allows us to access a scale where $\langle j(t)j\rangle$ is saturated, which enables a quantitative estimate of the Drude weight. This is illustrated in Fig.~\ref{fig:highT} for fixed $\Delta=0.6$ and three temperatures $T/J=0.5,2,\infty$. The current correlators become flat around $tJ\sim20$, and we obtain $D\sim0.06J/2T$ for all $T/J\gtrsim0.5$. Note that at $T/J=0.5$, the Drude weight is reduced by a factor of $3$ compared to its zero-temperature Bethe ansatz value \cite{betheT0}. We emphasize that for each temperature we chose a system size large enough to determine $D$ in the thermodynamic limit [this is demonstrated explicitly in the inset to Fig.~\ref{fig:dw}(b)].

It is instructive to compare our data with the bosonization calculation of Ref.~\onlinecite{sirkerJ2} which yields the current correlator up to the asymptotic value $D$. With our prediction for $D$, the bosonization curve at $T/J=0.5$ agrees qualitatively with the DMRG curve despite the large temperature (see the inset to Fig.~\ref{fig:highT}). This demonstrates that it is indeed reasonable to expect $\langle j(t)j\rangle$ to saturate at a scale $tJ\sim20$. At larger temperatures, the correlators evolve non-monotonously at intermediate $t$ \cite{sirkerJ1}. The magnitude of these oscillations, however, decreases as $T$ is lowered; they eventually die out around $T/J=0.5$.

We now investigate the $\Delta$--dependence of the Drude weight for intermediate ($T/J=0.5$) and high ($T=\infty)$ temperatures. As illustrated in Fig.~\ref{fig:dw}, $D$ monotonously decreases from its trivial value at $\Delta=0$ (where $j$ commutes with the Hamiltonian and thus $\langle j(t)j\rangle=\tn{const.}$) down to a finite value for the isotropic chain. The latter is consistent with most previous works which eventually concluded in favor of $D(\Delta=1)>0$ \cite{ed2,ed3,qmc1,bethekluemper,fieldtheory}. The qualitative behavior $D(\Delta_1)<D(\Delta_2)$ for $\Delta_1>\Delta_2$ agrees with one \cite{bethezotos} of the Bethe ansatz \cite{bethezotos,bethekluemper} calculations [note, however, that Ref.~\onlinecite{bethezotos} finds $D(\Delta=1)=0$ for any nonzero $T$]. At infinite temperature, our data is consistent with a recently-established lower Mazur bound \cite{mazurnonloc}. The latter has a fractal $\Delta$--dependence and seems to saturate our Drude weight for commensurate values $\Delta=\cos(\pi 
l/m), l,m\in\mathbb{Z}^+$ [such as $\Delta=0.5$; see the left inset to Fig.~\ref{fig:dw}(b)].

\begin{figure}[t]
\includegraphics[width=0.95\linewidth,clip]{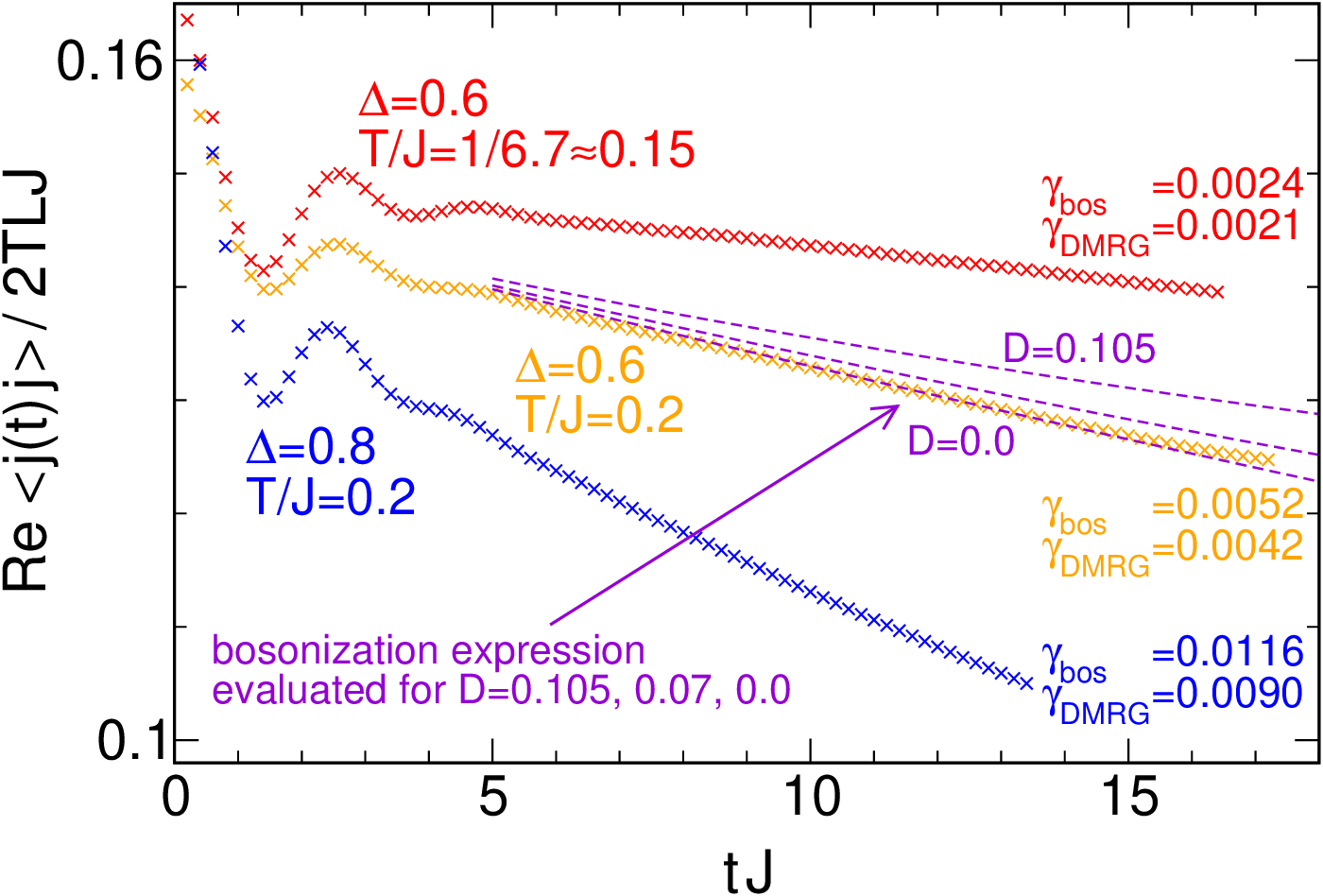}
\caption{Spin current correlation functions at low temperatures. Due to the presence of a diffusive transport channel \cite{sirkerJ1} with a small decay rate $\gamma$ (manifesting as a linear decrease of $\langle j(t)j\rangle$ for times $1/T\ll t\ll1/\gamma$), the Drude weight can only be estimated indirectly by comparing with the bosonization prediction \cite{sirkerJ2}. For $T/J=0.2$ and $\Delta/J=0.6$, this suggests $D\sim0.07$ as an upper bound. }
\label{fig:lowT}
\end{figure}

\emph{Low temperatures} --- At low $T$, bosonization \cite{sirkerJ1,sirkerJ2} conjectures a diffusive transport contribution: $\langle j(t)j\rangle$ falls off exponentially for times $t\gtrsim1/T$ and saturates at a scale set by the inverse decay rate $\gamma$ (bosonization does not yield the saturation value $\sim D$). This picture is strongly supported by DMRG \cite{sirkerJ1} as well as Quantum Monte Carlo \cite{qmc3} numerics. Our modified DMRG reconfirms the existence of the diffusive channel, and we extract $\gamma$ (by fitting the exponential in its linear regime $t\ll1/\gamma$) in nice agreement with the bosonization prediction (see Fig.~\ref{fig:lowT}; keep in mind that the DMRG calculation of Ref.~\onlinecite{sirkerJ1} reached $tJ\sim7$). Unfortunately, $\gamma$ decreases with temperature and for $T/J\lesssim0.4$ we can no longer access the scale $t\gg1/\gamma$ where the current correlator saturates.

We will now discuss an indirect way to determine $D$ which was proposed in Ref.~\onlinecite{sirkerJ2}. The only free parameter in this bosonization calculation for $\langle j(t)j\rangle$ is the Drude weight, which one can therefore try to extract by fitting the DMRG data at intermediate times. This is exemplified in Fig.~\ref{fig:lowT} for $\Delta=0.6$ and $T/J=0.2$. Our numerics \textit{suggests} $D\sim0.07$ as an upper bound (and seemingly contradicts both Bethe ansatz predictions $D=0.15$ \cite{bethekluemper} and $D=0.105$ \cite{bethezotos}). This in turn \textit{indicates} that the Drude weight only slightly deviates from its value $D\sim0.06$ at $T/J=0.5$ when temperature is lowered to $T/J=0.2$, and thus $D$ supposedly varies strongly below $T/J\lesssim0.2$ in order to eventually approach its zero-temperature value $D=0.15$ \cite{betheT0}. We emphasize, however, that these arguments are less stringent than the direct observation of a flattening $\langle j(t)j\rangle$: $T/J=0.2$ might be too high for 
the low-energy bosonization approach to be ultimately accurate -- but at lower $T$, we cannot access the time regime $1/T\lesssim t\lesssim 1/\gamma$ necessary for a meaningful comparison.


\emph{Gapped phase} --- The gapped phase of the Heisenberg chain has attracted less attention than the critical one: At $T=0$, the Drude weight vanishes for all $\Delta>1$ \cite{betheT0}, so one might reasonably expect the same to hold at finite temperature \cite{ed3,fabian,sirkerJ1,gaparg}. Indeed, our data for $T/J=0.5$ and $\Delta=1.2,1.4,1.6$ is consistent with a zero Drude weight [see Fig.~\ref{fig:dw}(a)]. However, the effort to reach large times increases significantly with $\Delta$, and we thus leave a detailed investigation of $D$ in the close vicinity of $\Delta\gtrsim1$ as a subject for future work.


\emph{Conclusion and Outlook} --- We have introduced a modification to finite-temperature DMRG which can be easily incorporated within existing implementations of the method. It allows to compute correlation functions up to significantly larger times. We have exploited this to quantitatively extract the Drude weight $D$ of the antiferromagnetic spin $1/2$ XXZ Heisenberg chain at intermediate to large temperatures. $D$ decreases monotonously when the asymmetry parameter $\Delta$ is varied from $\Delta=0$ (XX chain) to $\Delta=1$ (isotropic chain). Our data strongly suggests that $D$ is finite at $\Delta=1$. For small $T$, we reconfirmed the existence of a diffusive transport contribution in the critical phase. Our access to the Drude weight, however, is still only qualitative at low temperatures, and quantitatively bridging the gap between zero temperature (where one can carry out Bethe ansatz calculations) and our data at intermediate and large $T$ remains as the major open challenge. The DMRG algorithm 
presented may enable various applications -- investigation of spin and current correlation functions of non-integrable models being an obvious one.

\emph{Acknowledgments} --- We thank T.~Barthel and J.~Sirker for sending us their DMRG data, acknowledge discussions with T.~Prosen, and are grateful for support to the Deutsche Forschungsgemeinschaft via KA3360-1/1 (C.K.), DOE BES (J.H.B.), and the ARO OLE program (J.E.M.).



\begin{thebibliography}{*}
\bibitem{commentTaux}
More precisely: After each physical Trotter step $\Delta t$ we time-evolve the auxiliary sites using the same Hamiltonian (time-reversed and replacing physical with auxilliary sites) by a step $-\Delta t$.
\bibitem{commentchilocal}
This motivates an implementation of position-dependent matrix dimensions $\chi_n$ rather than working with a constant $\chi$. It turns out that while one can still efficiently limit the total discarded weight to a given bound $\epsilon$, the numerical speedup is significant.
\bibitem{commentnum}
In the unmodified DMRG $\chi$ increases so rapidly that extending the timescale by a factor of $3$ would supposedly require a computational effort of the order of weeks -- if possible at all.
\bibitem{discweight}
The DMRG calculation was carried out with a discarded weight chosen such that one obtains the current correlator with a relative error of $0.5-2\%$ ($3-4\%$ for the curve at $\Delta=0.2$, $T/J=0.5$).
\end{thebibliography}
\end{document}